\documentclass[%
 reprint,
superscriptaddress,
nofootinbib,
 amsmath,amssymb,
 aps,
]{revtex4-2}
\usepackage[utf8]{inputenc}
\usepackage{amsmath}
\usepackage[colorlinks=true]{hyperref} %color in references to fig,bib,etc
\hypersetup{colorlinks,
	citecolor=blue
}
\usepackage[dvipsnames]{xcolor}
\usepackage[caption=false]{subfig}
\usepackage{booktabs}

\usepackage{tikz}
\usetikzlibrary{patterns}
\usetikzlibrary{plotmarks}
\usetikzlibrary{matrix}
\usetikzlibrary{calc,shapes.misc,shapes.geometric}
\usepackage{slashed}
\usepackage{soul}

\usepackage{graphicx}
\usepackage{dcolumn}%
\usepackage{bm}
\newcommand{\beq} {\begin{equation}}
\newcommand{\eeq} {\end{equation}}
\newcommand{\beqn} {\begin{eqnarray}}
\newcommand{\eeqn} {\end{eqnarray}}
\begin{document}

\title{Relativistic distorted-wave analysis of the missing-energy spectrum measured with monochromatic $\nu_\mu$–$^{12}$C interactions at JSNS$^{2}$
}

\author{J. M. Franco-Patino}
\affiliation{
    INFN Sezione di Lecce, Via Arnesano, I-73100 Lecce, Italy
}

\author{J. García-Marcos}
\affiliation{
    Grupo de Física Nuclear, Departamento de Estructura de la Materia, Física Térmica y Electrónica, Facultad de Ciencias Físicas, Universidad Complutense de Madrid and IPARCOS, CEI Moncloa, Madrid, 28040, Spain
}
\affiliation{
    Department of Physics and Astronomy, Ghent University, B-9000 Gent, Belgium
}

\author{V. Belocchi}
\affiliation{Instituto de Física Corpuscular (IFIC), Consejo Superior de Investigaciones Científicas (CSIC) and Universidad de Valencia, E-46980 Paterna, Valencia, Spain}
\affiliation{School of Physics, The University of New South Wales (UNSW), Sydney, New South Wales 2052, Australia}
\affiliation{
    Dipartimento di Fisica, Universit\`a di Torino, Via P. Giuria 1, 10125 Torino, Italy
}
\affiliation{ 
    INFN, Sezione di Torino, Via Pietro Giuria 1, 10125 Torino, Italy
}

\author{M. B. Barbaro}
\affiliation{
    Dipartimento di Fisica, Universit\`a di Torino, Via P. Giuria 1, 10125 Torino, Italy
}
\affiliation{ 
    INFN, Sezione di Torino, Via Pietro Giuria 1, 10125 Torino, Italy
}

\author{G. Co'\,}
\affiliation{
   INFN Sezione di Lecce, Via Arnesano, I-73100 Lecce, Italy
}
\affiliation{
    Dipartimento di Matematica e Fisica ``E. De Giorgi'', Universit\`a del Salento, I-73100 Lecce, Italy
}

\author{R. González-Jiménez}
\affiliation{
    Departamento de Física Atómica, Molecular y Nuclear, Universidad de Sevilla, 41080 Sevilla, Spain
}

\date{\today}

\begin{abstract}
    Recently, the JSNS$^2$ collaboration measured for the first time the missing-energy distribution of $^{12}$C using a monochromatic neutrino beam coming from kaon decays at rest. In this work we present the results of an analysis of this spectrum using the relativistic distorted-wave approach. A new parameterization of the spectral function for neutrons in $^{12}$C, which incorporates detailed information from $\left(e,e'p\right)$ experiments with high missing-energy resolution has been used. The role of the recoil of the residual nucleus, final-state interactions, and neutrino event generators are discussed. 
\end{abstract}

\maketitle

\section{Introduction}
\label{sec:intro}  %-----------------------------------------%

Motivated by the global effort to measure the neutrino-oscillation parameters, neutrino–nucleus interactions have been investigated with unprecedented intensity~\cite{ALVAREZRUSO20181}. The interpretation of accelerator-based measurements requires accurate modeling of neutrino–nucleus cross sections over a wide range of energies. 
Since neutrino beams are commonly produced via charged-particle decays in flight, they exhibit a broad energy spectrum at the detector, which complicates the reconstruction of the incident neutrino energy on an event-by-event basis. Moreover, beyond the intrinsic uncertainties in energy reconstruction, determining the reaction mechanism underlying the measured observables remains a persistent and nontrivial challenge.

Kaon decays at rest (KDAR) provide a unique opportunity to study neutrino–nucleus interactions while avoiding the complications associated with neutrinos produced by meson decays in flight. The resulting monoenergetic $\nu_\mu$ beam, with $E_\nu = 235.5$ MeV, constitutes an especially clean probe for the direct study of nuclear effects such as initial- and final-state interactions and correlations in the nuclear medium. More broadly, KDAR $\nu_\mu$ offer the prospect of substantially reducing experimental and theoretical uncertainties and mitigating longstanding ambiguities.

Although KDAR neutrinos were observed for the first time by the MiniBooNE collaboration~\cite{PhysRevLett.120.141802}, the JSNS$^2$ collaboration \cite{mar25} has set a new paradigm in neutrino physics. For the first time, in a laboratory scattering experiment, the missing-energy spectrum of neutrons in $^{12}$C using KDAR neutrinos was measured. Unfortunately, only the shape of the cross section versus the missing energy is available and not its normalization. Despite this limitation, these first data represent an important step forward in the direction of understanding the low-energy neutrino-nucleus interaction.

In this article, we present the results of an investigation aimed to study the role played by various nuclear effects related to the neutrino-nucleus interaction. We start by using the relativistic-mean field model~\cite{RING1996193,WALECKA1974491,Serot1992} to describe the nucleus ground-state. We correct this basic model by adding effects beyond the mean-field approximation based on a parameterization of the $^{12}$C spectral function given in~\cite{PhysRevC.110.054612}. For the modeling of final-state interactions, we include proton distortion using the relativistic distorted-wave formalism~\cite{PhysRevC.48.2731,PhysRevC.51.3246,PhysRevC.64.024614,PhysRevC.105.025502,PhysRevC.105.054603,PhysRevD.106.113005, PhysRevD.109.013004} and simulate the inelastic interactions using the neutrino event generator NuWro~\cite{PhysRevC.86.015505,PhysRevC.100.015505}. 

In Sec.~\ref{sec:kinematics} we summarize the general kinematics of the interaction of a monochromatic neutrino with a nucleus in the impulse approximation. In Sec.~\ref{sec:model} and Sec.~\ref{sec:rho_par} we address the relativistic distorted-wave formalism, as well as the description of the ground-state nucleus within the relativistic-mean model with a parameterized spectral function. In Sec.~\ref{sec:results} we present a comparison of the JSNS$^2$ measurement and the theoretical predictions. Finally, we draw our conclusions in Sec.~\ref{sec:conclusions}. 

\section{Kinematics}\label{sec:kinematics} %-----------------------------------------%
We present in this section the basic expressions of our description of the 
process studied by the JSNS$^2$ collaboration where muon neutrinos
generated by the KDAR process impact on the $^{12}$C target inducing
a charge current (CC) reaction producing a $\mu^-$ of energy $E_\mu$. We work under the assumption that, given the low energy of the incoming neutrino, the main interaction channel is the quasielastic (QE) scattering off a single bound neutron in $^{12}$C. This means that the value of the energy transferred to the bound neutron $\omega = E_\nu - E_\mu$ is above the nucleon emission threshold. Therefore, our final state will be characterized by a residual nucleus $B$ and one proton emitted with energy $E_p$. The residual nucleus of mass $M_B$ has total energy $E_B$  which includes the recoil kinetic energy $T_B$ and an eventual  excitation energy $\xi$, such as $E_B = M_B + T_B + \xi$.

For convenience, from now on we choose to work in the laboratory frame, with the $z$ axis aligned with the incoming neutrino momentum $\mathbf{k_\nu}$. Within the impulse approximation (IA), where the exchanged $W$ boson is assumed to couple to a single bound nucleon that is subsequently ejected, the semi-inclusive $\nu_\mu + A \rightarrow \mu+p+B$ cross section can be written as~\cite{PhysRevD.106.113005}
\begin{widetext}
        \begin{equation}\label{Eq:1}
            \frac{d\sigma (E_\nu)}{dk_\mu d\Omega_\mu dp_pd\Omega_pdE_m} = \frac{G_F^2\cos^2{\theta_c}k_\mu^2p_p^2}{64\pi^5}L_{\mu\nu}H^{\mu\nu}\delta\left(\omega+M_A-E_p-E_B\right),
        \end{equation}
\end{widetext}
where $\mathbf{k}_\mu$ and $\mathbf{p}_p$ denote the momenta of the 
ejected muon and proton, respectively, and $\Omega_\mu$ and $\Omega_p$ 
represent their corresponding emission angles. We indicate with 
$M_A$ is the mass of the target nucleus, with $G_F$ 
the Fermi constant and with $\theta_c$ the Cabibbo angle. The cross section is also a function of the contraction of the leptonic $L_{\mu\nu}$ and the hadronic $H^{\mu\nu}$ tensors.

The observable measured by JSNS$^2$, the missing energy $E_m$, is defined as follows
    \begin{equation} \label{eq:menergy}
        E_m =  E_s^n + \xi = M_B-M_A+m_n + \xi,
    \end{equation}
where $E_s^n = M_B-M_A+m_n$ is the neutron separation energy. 
The energy conservation in Eq.~\eqref{Eq:1} can be expressed as
    \begin{equation}\label{Eq:2}
        \omega+M_A = E_p+E_B \longrightarrow \omega +m_n = E_m + T_p + m_p + T_B,
    \end{equation}
where $m_p$ is the mass of the emitted nucleon, $T_p$ its kinetic energy, and
$T_B$ is the kinetic energy of the residual nucleus
    \begin{equation}
        T_B = E_B-M_B -\xi = \sqrt{\mathbf{p}_m^2+(M_B+\xi)^2}-M_B -\xi.
    \end{equation}
We indicate with
 $\mathbf{p}_m = \mathbf{p}_p-\mathbf{q}$ the missing momentum, 
 which is fixed by momentum conservation and depends on the momentum 
 of the ejected nucleon $\mathbf{p}_p$ and on the transferred
 momentum $\mathbf{q} = \mathbf{k_\nu}-\mathbf{k_\mu}$. 

 The recoil of the residual nucleus $T_B$ in Eq.~\eqref{Eq:2} introduces 
 a complicated angular dependence in the energy-conservation delta function, 
 making the integration over one of the variables in Eq.~\eqref{Eq:1} non-trivial. 
 For instance, if we solve the energy conservation 
 to integrate on $p_p$ we obtain a quadratic equation
    \begin{equation}\label{Eq:conservation}
        \left[A_0^2 - \alpha^2\right]p_p^2 - \alpha\Delta p_p+A_0^2m_p^2-\Delta^2/4 = 0,
    \end{equation}
    where
    \begin{align}
        A_0&=E_\nu-E_\mu+M_A \\\nonumber
        \Delta&=A_0^2+m_p^2 -q^2-(M_B+\xi)^2 \\\nonumber
        \alpha= E_\nu\cos{\theta_p}-&k_\mu\left(\sin{\theta_\mu}\sin{\theta_p}\cos{\phi_p}+\cos{\theta_\mu}\cos{\theta_p}\right).
    \end{align}
Integrating over $p_p$ using the delta function introduces a factor 
\[
f_{\textnormal{rec}}^{-1} = \left|\frac{p_p}{E_p}+\frac{p_p-\alpha}{E_B}\right|^{-1}
\;\;,
\]
in the expression of the cross section. Then, Eq.~\eqref{Eq:1} becomes
    \begin{equation}\label{Eq:3}
        \frac{d\sigma (E_\nu)}{dk_\mu d\Omega_\mu d\Omega_p dE_m} =        \frac{G_F^2\cos^2{\theta_c}k_\mu^2p_p^2}  {64\pi^5f_{\textnormal{rec}}}L_{\mu\nu}H^{\mu\nu},
    \end{equation}
where $p_p$ is the solution of Eq.~\eqref{Eq:conservation}. If the two solutions of Eq.~\eqref{Eq:conservation} are valid, then the cross section is the incoherent sum of both solutions. As we will show later, retaining the recoil of the residual nucleus in Eq.~\eqref{Eq:2} allows us to investigate its impact on the missing-energy spectrum measured by JSNS$^2$.

%%%%%%%%%%%%%%
\section{The relativistic distorted-wave approach} \label{sec:model} 

The kinematic region accessible to the JSNS$^{2}$ experiment requires a careful 
treatment of nuclear effects (\textit{e.g.}  Pauli blocking, ground state correlations, re-interaction
of the emitted nucleon with the residual nucleus, etc.) in order to accurately describe
the experimental results~\cite{PhysRevC.103.064603}.  

In this work we adopt the relativistic distorted-wave approach 
based on the energy dependent-relativistic mean field model (ED-RMF)~\cite{PhysRevC.100.045501,PhysRevC.101.015503}.
This model has been used to describe inclusive $\left(e,e'\right)$ data 
that overlap in phase-space with 235.5 MeV $\nu_\mu$
scattering processes~\cite{PhysRevC.100.045501}.

In our description of the process we consider that $\nu_\mu$ interacts
with a bound neutron which becomes a proton because of the CC 
weak interaction. The proton is then emitted from the nucleus. 
In our model, the remaining nucleons act as inert spectators 
in the one-body process described above.
In the relativistic distorted-wave impulse approximation (RDWIA) we can express the hadronic tensor $H^{\mu\nu}$ 
of Eq.~\eqref{Eq:3} as a sum over occupied initial-states identified by the quantum number $\kappa$ 
    \begin{equation}\label{Eq:4}
       H^{\mu\nu} = \sum_\kappa \mathcal{H}^{\mu\nu}_{\kappa} = \sum_\kappa\rho_\kappa(E_m)\sum_{s_p, m_j}J^\mu_{s_p, \kappa, m_j} J^{\nu *}_{s_p, \kappa,m_j},
    \end{equation}   
where the hadronic current is 
    \begin{equation}
        J_{s_p,\kappa,m_j}^\mu = \int d\mathbf{r} e^{i\mathbf{r}\cdot\mathbf{q}}\bar{\Psi}^{s_p}(\mathbf{r},p_p)\hat{J}^\mu\Phi_\kappa^{m_j}(\mathbf{r}),
    \end{equation}
and $\rho_\kappa(E_m)$ indicates the missing energy profile.     
In the above expressions $j = |\kappa| - 1/2$ is the total angular momentum of the bound neutron,
$m_j$ its third component and $s_p$ is the spin projection of the ejected proton. 
The weak current one-body operator $\hat{J}^\mu$ has the CC2 expression 
given in Ref.~\cite{PhysRevD.106.113005}.
The wave functions $\Psi^{s_p}$ and $\Phi_\kappa^{m_j}$ are 
four-dimensional spinors that describe, respectively, the scattered 
and bound nucleons. 

\begin{figure}[t]
	   \centering
	   \includegraphics[width=0.49\textwidth]{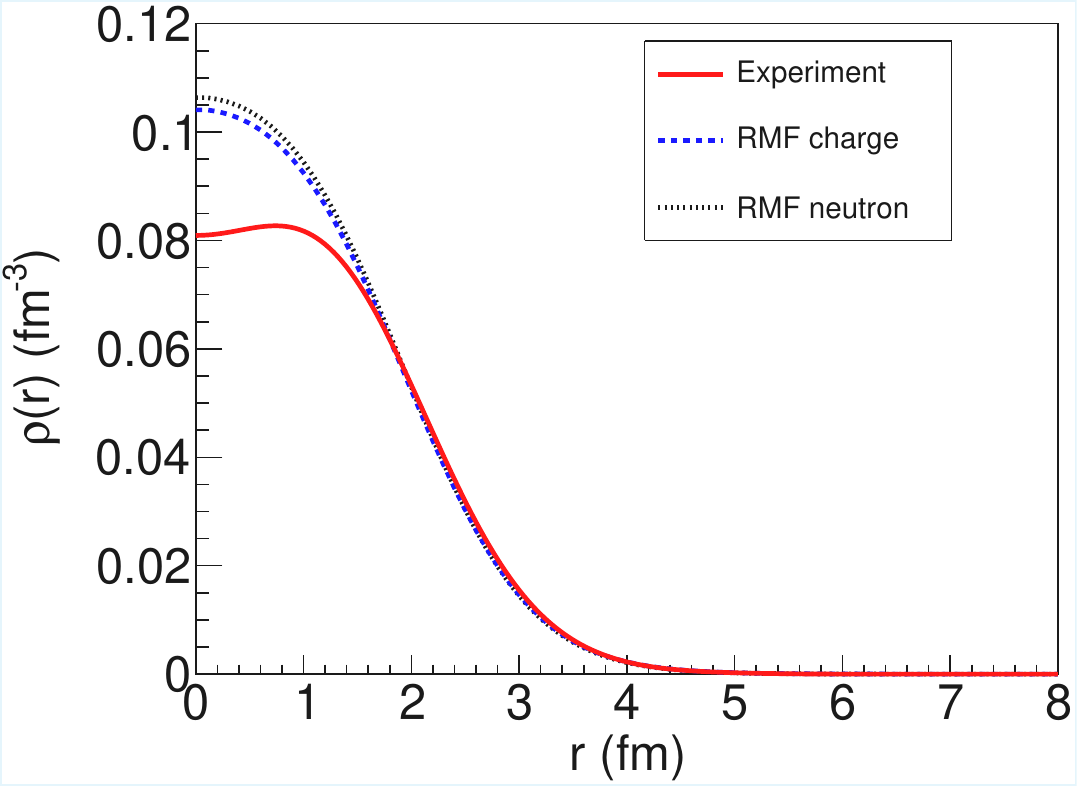}
      \caption{\label{fig:density} 
        Charge and neutron radial density calculated with the RMF potential NLSH~\cite{SHARMA1993377}, compared with a fit to elastic electron–nucleus scattering data~\cite{DEVRIES1987495}. The root-mean-square (rms) charge radius obtained with NLSH is $2.46811$ fm, in agreement with the experimental value of $2.4701 \pm 0.0022$ fm~\cite{ANGELI201369}.
        }
\end{figure}

The $^{12}$C ground state is described in terms of single particle 
wave functions $\Phi_\kappa^{m_j}(\mathbf{r})$ obtained within the independent-particle shell model, where the nucleons inside the nucleus occupy discrete energy states that are eigenvalues of a spherically symmetrical potential. For this work, we consider a fully relativistic framework where the bound states are solutions of the Dirac equation with scalar and vector potentials given by the relativistic mean field (RMF) approach~\cite{RING1996193,WALECKA1974491,Serot1992}. The free parameters of the RMF potential are fitted to reproduce different properties of finite nuclei. In this work, we use the potential NLSH described in~\cite{SHARMA1993377}. Table~\ref{tab:spenergies} lists the model predictions for the neutron and proton separation energies in $^{12}$C, while Fig.~\ref{fig:density} shows the corresponding densities and root-mean-square charge radius.

%grafica con la densidad y otras propiedades en una tabla.

%2.46811 proton RMS ANGELI201369, 2.4701 +- 0,0022 fm

 %{\color{blue}We indicate here the details of the potential used in the calculation. Which potential has been used. Results concerning the $^{12}$C ground state. Binding energy vs experimental one. Single particle energies of both proton and neutron states vs experimental values. one figure showing the charge distribution compared with the empirical one and a neutron density distribution.}{\color{red}This could be included, however the exact single-particle energies predicted by the RMF model itself aren't used at the end in the calculation because we introduce the parameterized profile. Instead we can reference NLSH paper https://doi.org/10.1016/0370-2693(93)90970-S that has comparison with some empirical data like binding energy, charge and neutron radii instead of adding the figure.}
 \begingroup
 	\setlength{\tabcolsep}{9.0pt}%adjust to the width of the column.
        \begin{table}[!h]
        	\centering
        	\begin{tabular}{ccccc}
        		\toprule\toprule
		       & \multicolumn{2} {c} {protons} & \multicolumn{2} {c} {neutrons} \\ \midrule
        		       & th. & exp. & th. & exp. \\ \midrule
         $1p_{3/2}$ & 14.46 &15.96 &17.81 & 18.72 \\ %\midrule
          $1s_{1/2}$ & 40.50 & - & 44.37 & - \\ %\midrule

        		\bottomrule\bottomrule
        	\end{tabular}
        	\caption{\label{tab:spenergies}
	Energy levels of the $1p_{3/2}$ and $1s_{1/2}$ shells obtained with the NLSH parameterization~\cite{SHARMA1993377} for protons and neutrons in $^{12}$C compared with the empirical separation energies from Ref.~\cite{Wang_2021}. All the energies are in MeV. }
        \end{table}
\endgroup
 
Within RDWIA, the wave function $\Psi^{s_p}\left(\mathbf{r}, p_p\right)$ describes the scattered proton with asymptotic momentum $p_p$ and is given by
\begin{widetext}
        \begin{align}
        	\Psi^{s_p}\left(\mathbf{r}, p_p\right) = 4\pi\sqrt{\frac{E_p+m_p}{2E_p}}\sum_{\kappa, \mu, m'}e^{i\delta_\kappa}i^l<l\,m'\,\frac{1}{2}\,s_p\,|\,j\,\mu>Y_{lm'}^{*}(\Omega_p)\psi_\kappa^\mu\left(\mathbf{r}, p_p\right),
        \end{align}
    \end{widetext}
where $Y_{lm}(\Omega_p)$ are the spherical harmonics 
that describe the direction of the emitted proton and $\delta_\kappa$ is the phase shift obtained by matching the distorted wave-function to the Dirac-Coulomb wave-function. The orbital momentum $l$ is fixed by $\kappa$ and take the values $l = \kappa $ if $\kappa > 0$ and $l = -\kappa - 1$ if $\kappa < 0$. The wave function $\psi_\kappa^\mu\left(\mathbf{r}, p_p\right)$ is a four-component spinor of the form
\begin{equation}
    \psi_\kappa^\mu\left(\mathbf{r}, p_p\right) = \begin{pmatrix}
    g_{\kappa}\!\left(E_p,r\right)\,\Phi_{\kappa}^{\mu}\!\left(\Omega_r\right)\\[4pt]
    i\,f_{\kappa}\!\left(E_p,r\right)\,\Phi_{-\kappa}^{\mu}\!\left(\Omega_r\right)\end{pmatrix}
\end{equation}
with the Pauli spin-spherical harmonics
\begin{equation}
    \Phi_{\kappa}^{\mu}(\Omega_r) = \sum_{s_p} <l \,m_l\,\frac{1}{2}\,s_p\,| j\,\mu>Y_l^\mu(\Omega_r)\chi^{s_p},
\end{equation}
where $\chi^{s_p}$ is a Pauli spinor of two components and $g_\kappa$ and $f_\kappa$ satisfy the following coupled-differential equations
\begin{align}
    \frac{df_\kappa}{dr} = &\frac{\kappa-1}{r}f_\kappa - \bigg(E_p-m_p-V(r)\bigg)g_\kappa \\\nonumber
    \frac{dg_\kappa}{dr} = -&\frac{\kappa-1}{r}g_\kappa + \bigg(E_p+m_p-V(r)\bigg)f_\kappa.
\end{align}
Our theoretical predictions for the QE channel correspond to two calculations based on the distorted-wave approach with two different potentials $V(r)$. One of the models is based on the ED-RMF potential~\cite{PhysRevC.100.045501,PhysRevC.101.015503}, which is real and used to describe inclusive $\left( e,e'\right)$ and $\left(\nu,l\right)$ data, and the other one is based on the relativistic optical potential (ROP)~\cite{PhysRevC.47.297}, which is complex and can be used to describe $\left(e,e'p\right)$ data~\cite{PhysRevC.48.2731,PhysRevC.51.3246,PhysRevC.64.024614}. More specifically, we use here the energy-dependent A-independent Carbon (EDAI-C) optical potential.  Notice that there is a third option, which is not realistic for the kinematics involved in the process, that is the relativistic plane-wave model (RPWIA) where the coupled equations are solved with $V(r) = 0$~\cite{PhysRevC.100.045501}.

%

%%%%%%%%%%%%%%%%%%
\section{Missing energy profile parameterization}\label{sec:rho_par} %-----------------------------------------%

In the pure shell model presented so far, the missing-energy distribution 
$\rho_\kappa\left(E_m\right)$ is given by a sum of Dirac delta functions. 
These delta functions ensure that the missing energy can take 
only discrete values, corresponding to the energy eigenvalues of the fully 
occupied single-particle states given the RMF model.
This model is relatively simple, 
and a meaningful comparison with the missing-energy distribution measured by JSNS$^2$ requires a more realistic description 
of the initial nuclear bound state that incorporates correlations, 
{\sl i.e.}, effects beyond the pure mean-field approximation.

There is an extensive body of literature addressing microscopic 
calculations of spectral functions, which lead to the missing-energy 
profile $\rho_\kappa\left(E_m\right)$~\cite{PhysRevC.49.R17,BENHAR1994493,PhysRevD.72.053005,PhysRevC.55.810,ATKINSON2019135027,Barbieri2017,SARTOR197629,JEUKENNE197683}.
Here, we adopt a phenomenological approach and, based on previous 
work~\cite{PhysRevC.105.025502,PhysRevD.106.113005}, we introduce 
a new parameterization of the missing-energy profile 
$\rho_\kappa\left(E_m\right)$ for neutrons in $^{12}$C. 
The parameterization is based on the updated spectral 
function of Ref.~\cite{PhysRevC.110.054612} 
which is optimized to improve the agreement with exclusive $\left(e,e'p\right)$ data~\cite{VANDERSTEENHOVEN1988547}. 
The improvement is obtained by modeling 
the region dominated by the $p$-states 
with three Dirac deltas instead of only one narrow Gaussian. 

In our modeling of $\rho_\kappa\left(E_m\right)$, three distinct 
contributions are considered. 
The first contribution is associated with the occupancy of the 
$p$-states and corresponds to the energy region extending 
from the neutron emission threshold to about
22 MeV ($E_m \approx $ 18-22 MeV). In this region, the profile is represented by three Dirac delta functions as in Ref.~\cite{PhysRevC.110.054612}. Those three p-states correspond to a $1p_{3/2}$ shell describing the ground state and two excited states, namely $1p_{1/2}$ and $1p_{3/2}$, corresponding to $2.125$ MeV and $5.020$ MeV of excitation energy~\cite{VANDERSTEENHOVEN1988547}. For both excited states we use the same wave-function predicted by the RMF model for the $1p_{3/2}$ ground-state.
%
%%%%%%%%
\begin{figure}[!htbp]
	   \centering
	   \includegraphics[width=0.49\textwidth]{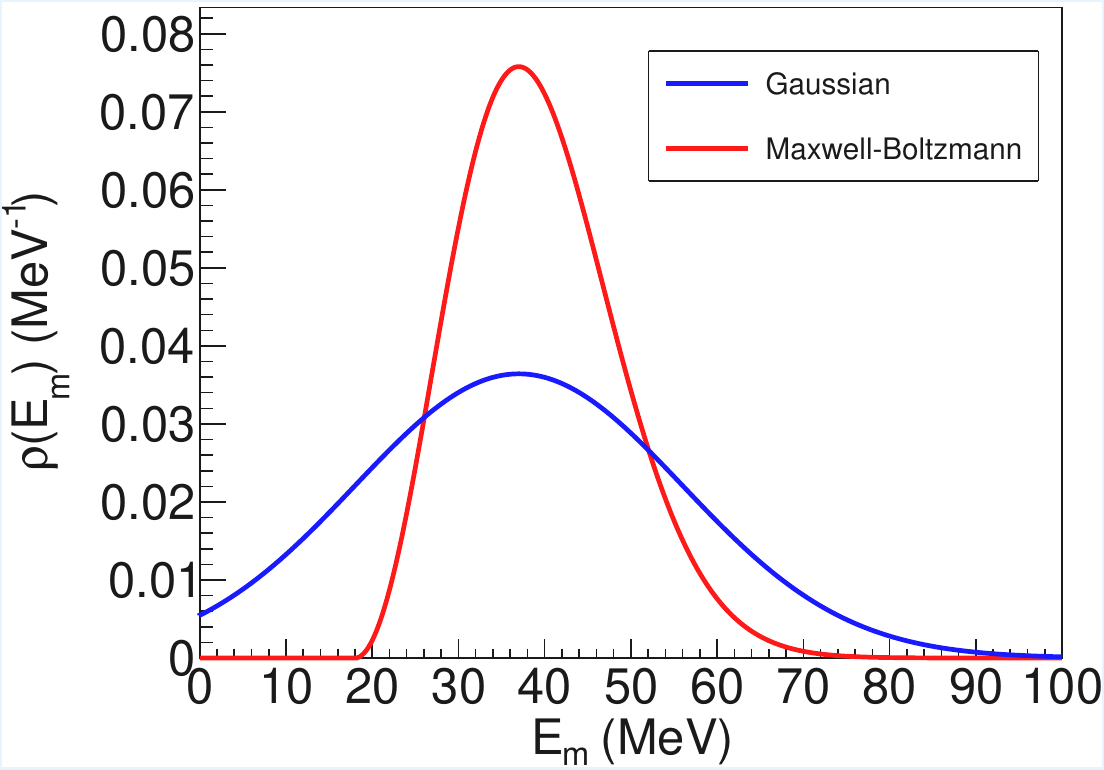}
      \caption{\label{fig:gauss_vs_mb} 
      Missing energy profile of a neutron in a $s$-state of $^{12}$C parameterized 
      with $\mu=37$ MeV and $\sigma=19$ MeV using Gaussian 
      and Maxwell-Boltzmann distributions (see Eq.~\eqref{eq:M-B}). 
      Both distributions are normalized to $n_{1s1/2} =$ 1.74 nucleons.}
    \end{figure}
%%%%%

The second contribution is associated with the $s$-states, which 
$(e,e'p)$ experiments indicate to exhibit a large width. We 
abandoned the usual gaussian parameterization
in favor of an asymmetric Maxwell-Boltzmann shape of the type
%
    %\begin{widetext}   
    
        \begin{equation} \label{eq:M-B}
        \resizebox{\columnwidth}{!}{$
            \rho_{1s1/2}\left(E_m\right) = \frac{4{n_{1s1/2}}}{\sigma\,\sqrt{\pi}}\;
            \exp\!\left(-\frac{(E_m-\mu+\sigma)^{2}}{\sigma^{2}}\right)\;
            \frac{(E_m-\mu+\sigma)^{2}}{{\sigma}^{2}} ;
        $}
        \end{equation}
    %\end{widetext}
%
with $\mu= 37.0$ MeV, $\sigma=$19 MeV and where the condition $E_m \ge \mu -\sigma$ is imposed. In Fig.~\ref{fig:gauss_vs_mb} we compare the missing-energy profiles 
of a neutron in a $s$-state  obtained using Gaussian and  Maxwell-Boltzmann 
distributions. The two distributions are normalized at the same value, 
corresponding to $n_{1s1/2} =$ 1.74 nucleons.
The Maxwell-Boltzmann distribution avoids the unphysical contribution
at $E_m$ values below the nucleon emission threshold. 

%%%%%%%%%%%%%
\begin{figure}[!htbp]
	   \centering
	   \includegraphics[width=0.49\textwidth]{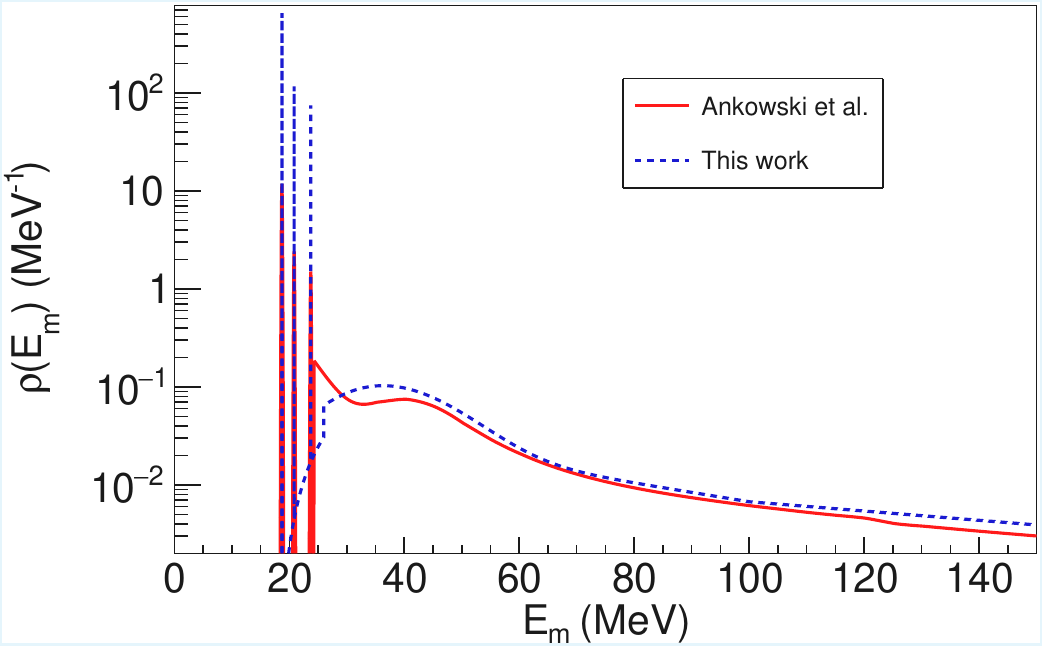}
      \caption{\label{fig:rho} 
Comparison of our missing energy profile $\rho_\kappa\left(E_m\right)$ 
of neutrons  with that of Ref.~\cite{PhysRevC.110.054612}. 
A shift of +2.764 MeV has been applied to the results of
Ankowski \textit{et al.} to account for neutron, rather than proton, emission
(see text for explanation).
}
\end{figure}
%%%%%
The third contribution in our modeling of $\rho_\kappa\left(E_m\right)$ is related to the region of $E_m$ and $p_m$ where the effects of short-range correlations on the spectral function are significant. This contribution would be strictly zero in the absence of NN correlations and its leading contribution comes from $2h1p$ states~\cite{BENHAR1994493}. Using this model, we include in an effective way the contribution of one nucleon emission from the correlated NN pairs in the nucleus. Although for large $p_m$ and $E_m$ there must be certainly emission of two nucleons, in the short-range correlation part of the spectral function we only consider that one nucleon is ejected. We refer to the contribution of two bound nucleons in a correlated pair to one-nucleon knockout as the “background,” and parametrize it as follows in order to reproduce the high-energy tail reported in~\cite{PhysRevC.110.054612}:
%%%
    \begin{equation}
        \rho_{\textnormal{backg}}\left(E_m\right) = n_b \exp(-x_{\textnormal{cut}}b)\exp\bigg(-w\big(E_m-x_{\textnormal{cut}}\big)\bigg)
    \end{equation}
    for $ 26 < E_m < 100 $ MeV, and
    \begin{equation}
        \rho_{\textnormal{backg}}\left(E_m\right) = n_b\exp(-b E_m)
    \end{equation}
 for $ 100 < E_m < 300$ MeV. The parameters values are $n_b = 0.02$, $b = 0.0109$ MeV$^{-1}$, $w = 0.022$ MeV$^{-1}$ and $x_{\textnormal{cut}} = 100$ MeV.
 
Our model for  $\rho_\kappa\left(E_m\right)$
is obtained by summing these three contributions, 
each normalized to the occupancy values listed in Table~\ref{Table:1}. 
The Fig.~\ref{fig:rho} shows a comparison of our $\rho_\kappa\left(E_m\right)$
with that of Ref.~\cite{PhysRevC.110.054612}. The main difference between
the two profiles occurs in the region where the contributions of $p-$ and $s-$shells overlap.
This explains the discrepancies between our occupation values and those reported in the
 $\left(e,e'p\right)$ experiments of 
Refs.~\cite{VANDERSTEENHOVEN1988547,MOUGEY1976461}, as
reproduced by Ankowski \textit{et al.}~\cite{PhysRevC.110.054612}.
Also, the distribution has been shifted by +2.764 MeV to account 
for the fact that we consider neutron emission, 
whereas Ref.~\cite{PhysRevC.110.054612} refers to proton emission. 
\begingroup
\setlength{\tabcolsep}{9.0pt}%adjust to the width of the column.
        \begin{table}[!h]
        	\centering
        	\begin{tabular}{ccc}
        		\toprule\toprule
        		                  & $E_\kappa$ (MeV) &$n_\kappa $ \\ \midrule
        		$1p_{3/2}$ & 18.724  &1.95\\
                $1p_{1/2}$ & 20.844  &0.29\\
                $1p_{3/2}$ & 23.734  &0.22\\
                $1s_{1/2}$ & 37.0   &1.74\\
                background & - & 1.80 \\
        		\bottomrule\bottomrule
        	\end{tabular}
        	\caption{\label{Table:1} 
	Energies $E_\kappa$ and neutrons number $n_k$
	for the different contributions in our modeling of
	 $\rho_\kappa\left(E_m\right)$. In the case of
	 the $s$-state, $E_\kappa$ denotes the peak position of the Maxwell-Boltzmann
	ditribution.
	}
        \end{table}
\endgroup

% 
% 
%    \begingroup
%		\setlength{\tabcolsep}{9.0pt}%adjust to the width of the column.
%        \begin{table}[!h]
%        	\centering
%        	\begin{tabular}{cccccccc}
 %       		\toprule\toprule
 %       		$\kappa$ & &$E_\kappa$ (MeV) & &$\sigma_\kappa$ (MeV) & &$n_\kappa $ \\\midrule
 %       		$1p_{3/2}$ & &18.724 & & - & &1.95\\\midrule		
 %               $1p_{1/2}$ & &20.844 & & - & &0.29\\\midrule	
%                $1p_{3/2}$ & &23.734 & & - & &0.22\\\midrule	
%                $1s_{1/2}$ & &37.0 & & 19 & &1.74\\\midrule
%        		\bottomrule\bottomrule
%        	\end{tabular}
%        	\caption{\label{Table:1} Parameterization of the missing energy profile with contributions from three p-shells and one s-shell. The parameter $E_\kappa$ is the position of the shell, $\sigma_\kappa$ the width and $n_\kappa$ the number of nucleons in each shell. Notice that the p-shells do not have any width because they are modeled as Dirac deltas. The missing 1.80 nucleons are allocated to the background.}
%        \end{table}
%    \endgroup
%%%%%%%%%%%    
\section{Comparison with experimental data}\label{sec:results} %-----------------------------------------%

The cross-section data reported by the JSNS$^{2}$ 
collaboration~\cite{mar25} is presented as a function of the missing energy $E_m$. 
This quantity, defined in Eq.~\eqref{eq:menergy}, is reconstructed experimentally using the following expression
\beq \label{eq:emiss1}
E_m= E_\nu - m_\mu + m_n-m_p - E_{\textnormal{vis}}
        \;\;,
\eeq
where $E_{\textnormal{vis}}$ is the visible energy collected by JSNS$^2$ 
liquid scintillator detector.
In the pure QE regime, where only a single proton is emitted,  $E_{\textnormal{vis}}$ 
can be obtained using Eq.~\eqref{Eq:2}
\beq \label{eq:evis}
E_{\rm vis} = T_\mu + T_p + T_B
\;\;. 
\eeq
This represents a simplified description of the process. Additional effects may modify the value of 
$E_{\textnormal{vis}}$ and, consequently, of $E_m$;
for example, photons emitted during nuclear de-excitation 
or neutron emission induced by the re-scattering of the outgoing proton.

\begin{figure}[!htbp]
	   \centering
	   \includegraphics[width=0.49\textwidth]{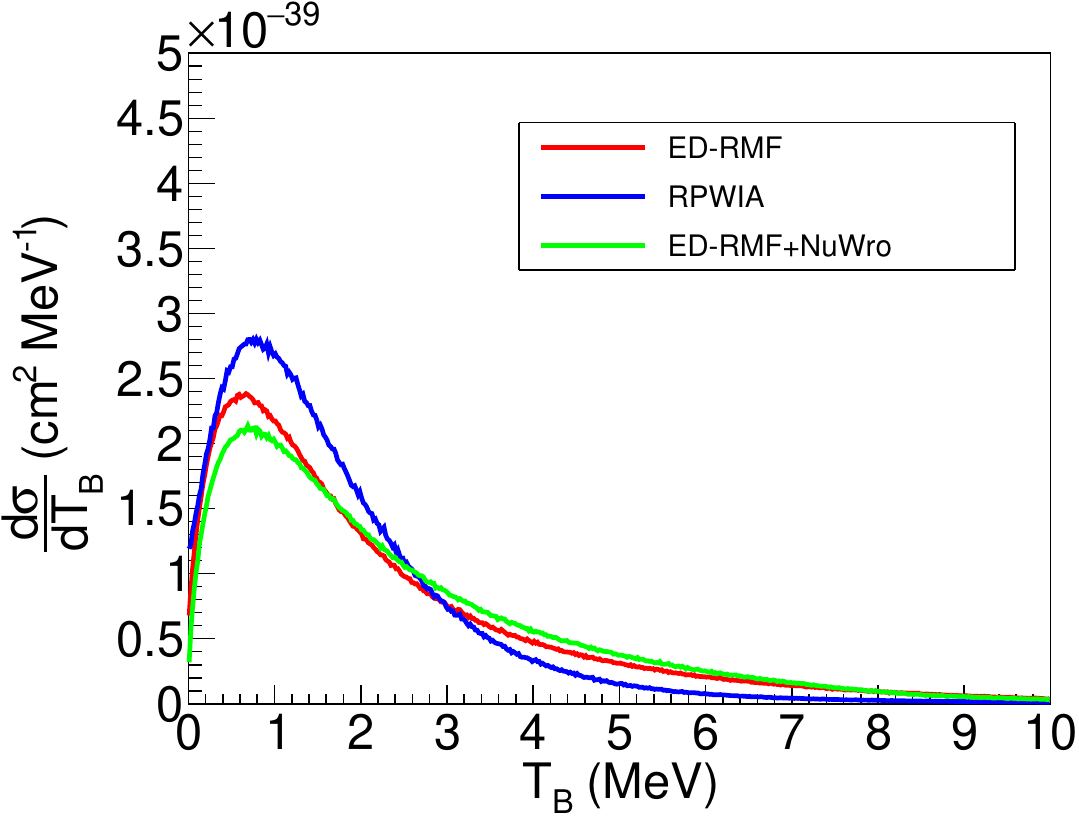}
      \caption{\label{fig:Tb_comparison} 
       Cross section as function of the kinetic energy of the residual $^{11}$C nucleus predicted by the RPWIA, ED-RMF and ED-RMF+NuWro models for the JSNS$^2$ kinematics (CC $\nu_\mu-^{12}$C interaction with $E_\nu = 235.5$ MeV).
        }
\end{figure}

We emphasize here an important issue regarding the definition of $E_{\rm vis}$. In~\cite{mar25} the JSNS$^{2}$ collaboration states that the experimental signal include small contributions ($\approx 1$ MeV) from nuclear recoil energy, photon emission from nuclear de-excitation or neutron-induced energy deposition because they are difficult to model using neutrino event generators. Since our formalism explicitly takes into account the nuclear recoil, in Fig.~\ref{fig:Tb_comparison} we show the predictions by the RPWIA and EDRMF models of the differential cross section as function of the residual nucleus recoil. The distorted-wave calculation exhibits a larger cross section in the tail of the distribution compared to the plane-wave prediction, effect that has been already found in $(e,e'p)$ analyses in the past~\cite{PhysRevC.51.3246,CABALLERO1998323}. As one could expect, the maxima are localized at a small value of $T_B \approx 0.8$ MeV, however the distributions has a large tail that extends beyond 4-5 MeV. This is compatible with the fact that the RMF wave-function of neutrons inside $^{12}$C has non-zero contribution for $p_m$ above the Fermi momentum $k_F = 225$ MeV~\cite{PhysRevD.104.073008}. For instance, for a bound neutron in the tail of the momentum distribution with $p_m = 320$ MeV, the recoil can be estimated to be $T_B \approx p_m^2/2M_B \approx$ 5 MeV for a $^{11}$C residual nucleus with mass $M_B \approx 10.25$ GeV. 
The results in Fig.~\ref{fig:Tb_comparison} show that the contribution from the nuclear recoil could be larger than JSNS$^2$ suggests. Also, one may question whether the kinetic energy of the residual nucleus can be fully or partially converted into detectable visible energy in the detector response. Even if the residual nucleus deposits kinetic energy in the scintillator, the associated light yield may be partially quenched or only partially reconstructed. In that case, part of the recoil contribution could effectively appear as additional missing energy. Since this issue remains open, in the following we present theoretical predictions both including and neglecting the recoil contribution when comparing with the experimental data. In this way, we quantify the impact of omitting the recoil term from $E_{\rm vis}$.

Another important issue concerns neutron emission. 
This process may occur when the primary proton produced at the interaction
vertex by the neutrino undergoes re-scattering inside the nucleus, 
interacting with a neutron so that both nucleons are emitted. 
Since neutrons do not generate detectable signals in the detector, the measured value of
$E_{\rm vis}$ in Eq.~\eqref{eq:evis} is smaller than that implied by energy conservation.
As a result, the reconstructed value of $E_m$ in Eq.~\eqref{eq:emiss1} is
overestimated with respect to its true value. Consequently, events in a $E_m$ bin are shifted toward higher missing-energy bins.

Our RDWIA approach takes into account the distortion of the scattered 
proton in a quantum-mechanical way, however neither the ED-RMF 
nor the ROP models are able to simulate the neutron emission.
In the ROP approach many effects beyond the single proton emission
are effectively considered in the imaginary part of the potential, but it is not possible to identify the neutron 
emission contribution. 

We tackle this problem by employing one-proton emission events generated according to the ED-RMF model as input to a generator's semiclassical cascade. The cascade subsequently redistributes the initial $\left(1p\right)$ strength into more complex final states, including $\left(1p1n\right)$, $\left(1pNn\right)$, $\left(2p\right)$, etc. This method has been applied~\cite{PhysRevC.105.054603,PhysRevC.110.054611,f7x5-snmz,Garcia-Marcos:2025dgl} to benchmark cascade models against ROP predictions, as well as to perform studies and comparisons with other neutrino-nucleus scattering data. In this work, we employ the NuWro~\cite{PhysRevC.86.015505,PhysRevC.100.015505} intranuclear cascade model to evaluate the effects of inelastic neutron emission in the comparison with the JSNS$^2$ missing-energy measurement. To avoid double counting the distortion of the proton by the residual nucleus, we make sure that NuWro only includes inelastic FSI. Any elastic FSI introduced by NuWro is discarded, since it is already taken into account by the ED-RMF model.

    \begin{figure}[!htbp]
	   \centering
	   \includegraphics[width=0.49\textwidth]{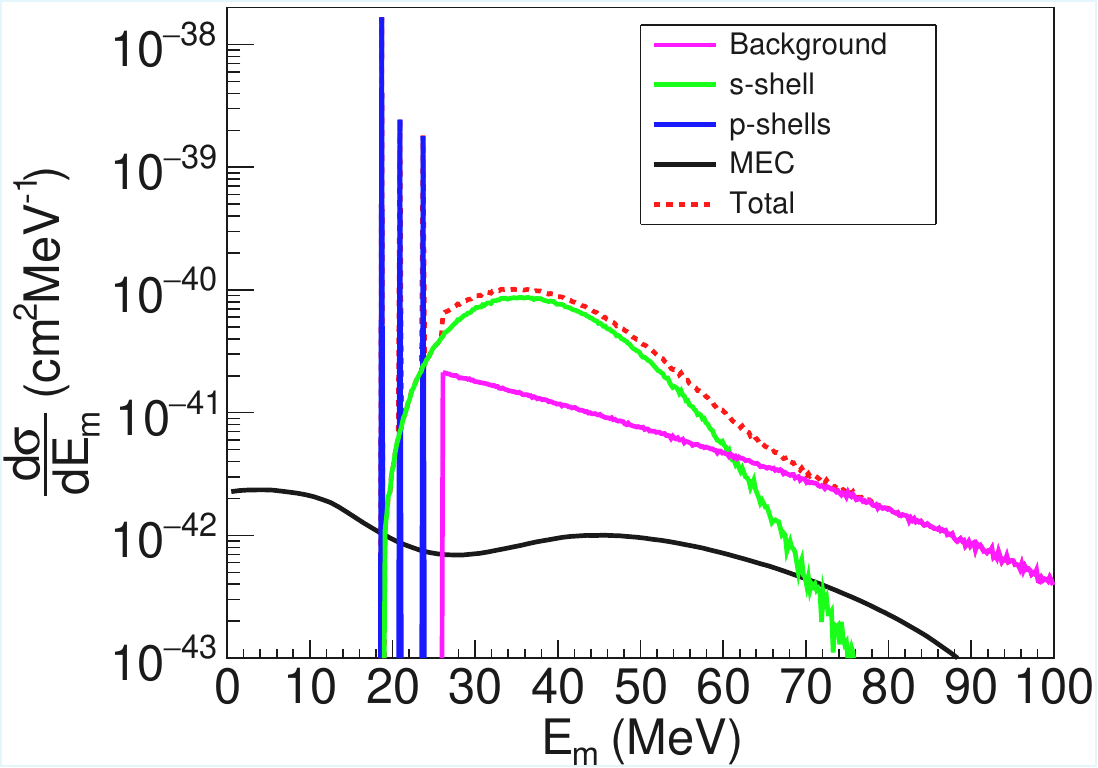}
      \caption{ \label{fig:by_shells} Differential cross section as function of the missing energy separated by contributions of the different shells to the QE channel using the ED-RMF model with recoil. The contribution of MEC calculated with a semi-inclusive model based on the Relativistic Fermi gas~\cite{Belocchi:2024rfp,Belocchi:2025eix} is also shown separately.}
    \end{figure}
%%%%%%%
    %%%%%%%%%%%%%    
    \begingroup
    \setlength{\tabcolsep}{1.0pt}
    \begin{table}[!h]
    	\centering
        \begin{tabular}{lcccc}
        \hline
            \noalign{\vskip 1mm} % space after the header line (adjust)
            Model & \multicolumn{4}{c}{Integrated cross-sections (cm$^2$)} \\
            \noalign{\vskip 1mm} % space after the header line (adjust)
            \cline{2-5}
            \noalign{\vskip 1mm} % space after the header line (adjust)
             & $p$-shells & $s$-shell & Background & Total \\
             \noalign{\vskip 1mm} % space after the header line (adjust)

            \hline
            \noalign{\vskip 1mm} % space after the header line (adjust)
            RPWIA  & $5.06\times10^{-39}$ & $2.87\times10^{-39}$ & $6.6\times10^{-40}$ & $8.59\times10^{-39}$ \\
            ED-RMF & $4.18\times10^{-39}$ & $1.89\times10^{-39}$ & $4.6\times10^{-40}$ & $6.53\times10^{-39}$ \\
            ROP   & $3.02\times10^{-39}$ & $1.39\times10^{-39}$ & $3.6\times10^{-40}$ & $4.77\times10^{-39}$ \\
            MEC   & - & - & - & $1.01\times10^{-40}$ \\

            \hline
        \end{tabular}
    	\caption{\label{Table:2} Total integrated cross section predicted by the different RDWIA QE models and MEC. For the QE models, the contributions are separated by shells.}
    \end{table}
    \endgroup
%    
%%%%%%%%%%%%%%%
%

We show in Fig.~\ref{fig:by_shells} the cross section obtained from ED-RMF calculation, which
considers the nuclear recoil. The different terms contributing to this results are separately shown.
As illustrated in Fig.~\ref{fig:rho} and in Table~\ref{Table:1}, 
the largest contribution in the $p$-shell region arises from the $1p_{3/2}$ state with 
a missing energy corresponding to the neutron separation energy of 
$^{12}$C, namely $E_m \approx 18.7$ MeV. In contrast, the contribution of the neutrons from the $s$ state is much broader, peaking at $E_m \approx 37.0$ MeV. The contribution of meson-exchange currents (MEC) is also shown separately in Fig.~\ref{fig:by_shells}. This process corresponds to the emission of two nucleons induced by the coupling of the $W$-boson to the two-body MEC in the primary vertex. Its contribution to the cross section is calculated using a semi-inclusive model based on the Relativistic Fermi gas (RFG)~\cite{Belocchi:2024rfp,Belocchi:2025eix} with two parameters, the Fermi momentum $k_F = 225$ MeV and an energy shift $E_{\textnormal{shift}} = 40$ MeV. The latter is introduced to account for the nucleons binding energy and final-state interactions within the RFG model. The MEC contribution shown in Fig.~\ref{fig:by_shells} is different from zero even below the one-nucleon emission threshold, a consequence of the oversimplified nuclear model used for this channel and the phenomenological parameter $E_{\textnormal{shift}}$ introduced to take FSI into account in this channel. The background and MEC contributions are significantly smaller and exhibit a broader distribution.

The relative importance of the various contributions can be more clearly assessed by examining the results reported 
in Table~\ref{Table:2}, where their integrated values are presented. The dominant contribution is that of the $p$-shell and that of the $s$ is about half of it. The contribution of the background is one order of magnitude smaller, and even smaller is the contribution of MEC. 

%%%%%%%%%%%%
    %\captionsetup[figure]{labelformat=simple,position=bottom}
    \begin{figure}[!htbp]
    	\captionsetup[subfigure]{labelformat=parens, position=top} 
    	\centering
    	{
    		\subfloat[]{\includegraphics[width=0.49\textwidth]{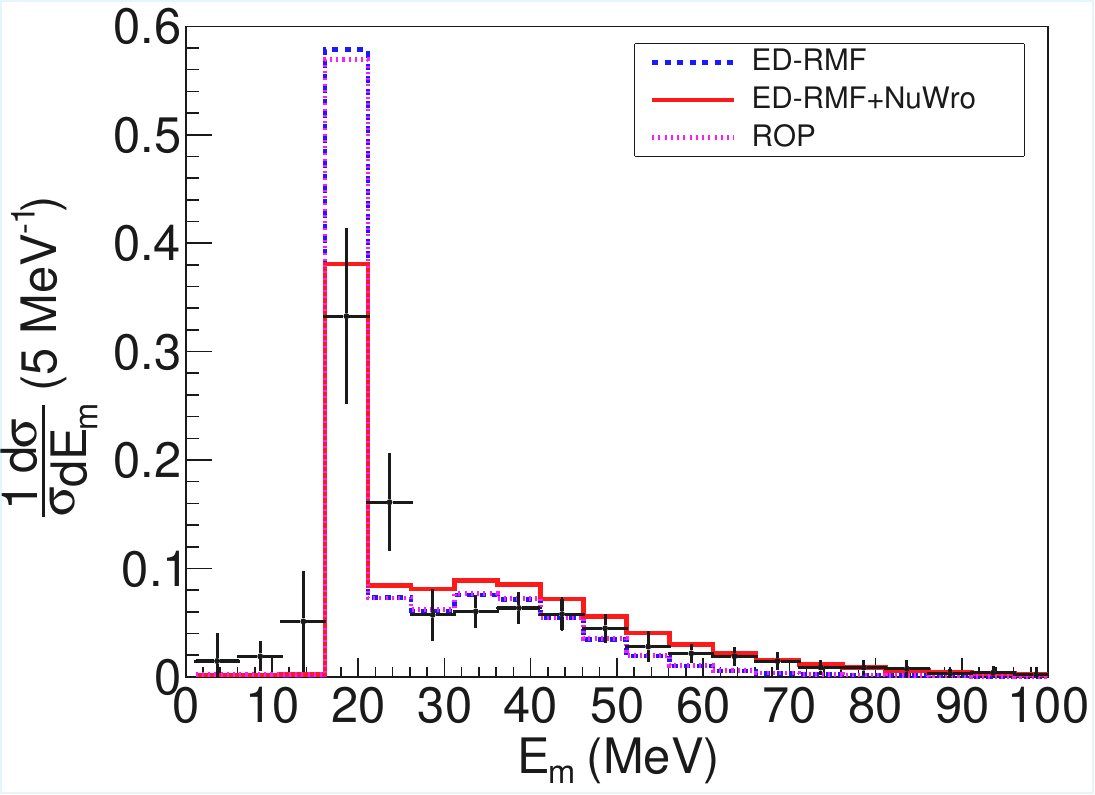}}
    		\:
    		\subfloat[]{\includegraphics[width=0.49\textwidth]{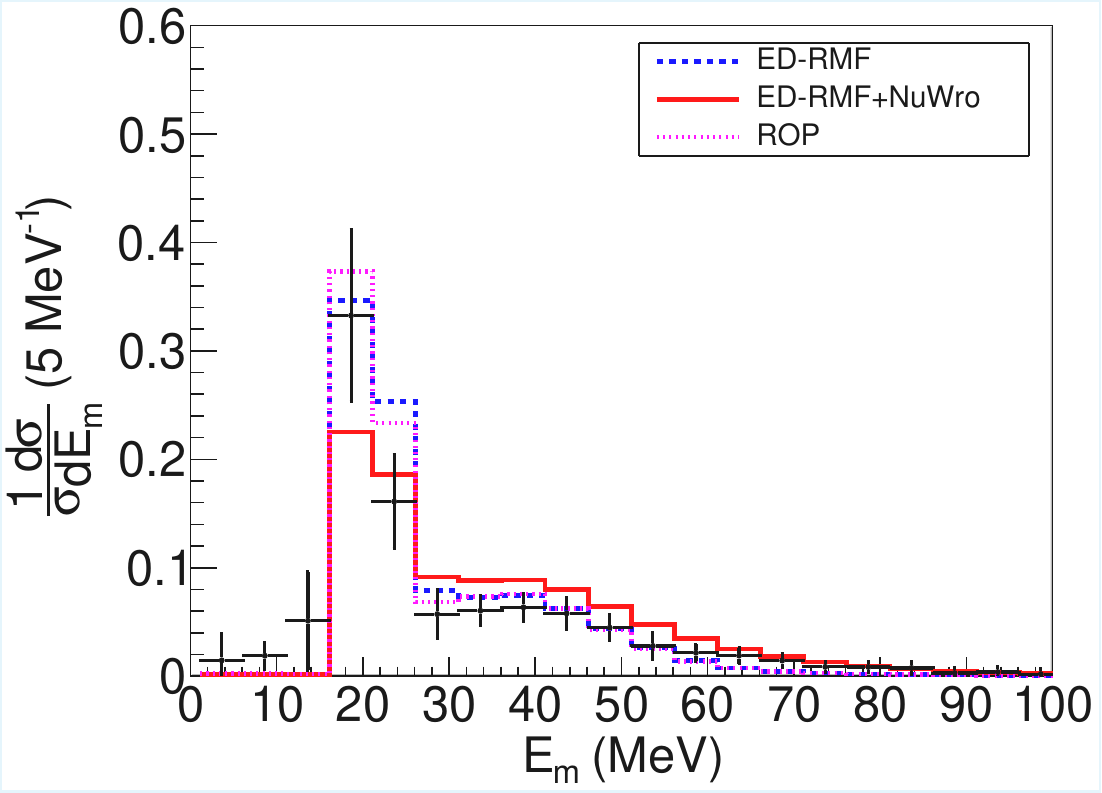}}
    	}
    	\caption{\label{fig:not_smeared} 
        Charge-current $\nu_\mu + ^{12}$C cross sections as a function of the missing energy $E_m$ for an incoming neutrino with $E_\nu = 235.5$ MeV.
        All the theoretical results are normalized to the total strength of the experimental data.
        In the panel~(a), the theoretical results consider the recoil of the residual nucleus
        in the definition of  $E_m$. In the panel~(b), this effect is neglected. All the results include both QE and MEC contributions.}
    \end{figure}
%%%%%%%%%%   

The comparison with the JSNS$^2$ experimental data requires two types of adjustments to our theoretical results. First, our cross section must be divided into bins of $E_m$. Second, a normalization procedure is necessary, since the JSNS$^2$ experiment provides only the shape of the cross section and not its absolute value.

The cross sections obtained after this procedure, calculated with the ED-RMF and ROP models, together with that obtained by implementing the ED-RMF results with the cascade simulation of NuWro, are shown in Fig.~\ref{fig:not_smeared}. The results of the upper panel include the nuclear recoil which is not considered in those of the lower panel.

In Fig.~\ref{fig:not_smeared} the results of the ED-RMF and ROP calculations are very similar. 
This similarity arises from the common normalization applied to both results. Table~\ref{Table:2} presents the total contributions from the two calculations, showing that the total cross section predicted by the ROP model is about 27\% smaller than that of the ED-RMF model. This is understandable, since the optical potential reduces the flux in the primary proton-emission channel, effectively redistributing it among other final-state channels.

The comparison between the results shown in the two panels clarifies the role of the residual-nucleus recoil. As already pointed out, neglecting this term in the evaluation of $E_{\rm vis}$ (see Eq.~\eqref{eq:evis}) leads to an overestimation of the missing energy. In our calculations, this corresponds to a shift of approximately 1 to 5 MeV. As a consequence, when the recoil is not included, a large part of the $p$-shell strength appears in the $20 < E_m < 25$ MeV bin, as shown in the lower panel. As shown in Fig.~\ref{fig:Tb_comparison}, the NuWro intranuclear cascade slightly modifies the residual nucleus recoil distribution predicted by the ED-RMF model. Changes in the number of nucleons in the final state and their kinematics introduced by NuWro reduce the peak of the $T_B$ distribution and increase the tail; therefore, shifts of up to 5 MeV in missing energy due to the recoil can still be obtained after the NuWro cascade is applied to the ED-RMF events.
%%%%%%%%%%%%%%%%%%%%%%%
    \begin{figure}[!htbp]
        \captionsetup[subfigure]{labelformat=parens, position=top} 
    	\centering
    	{
    		\subfloat[]{{\includegraphics[width=0.49\textwidth]{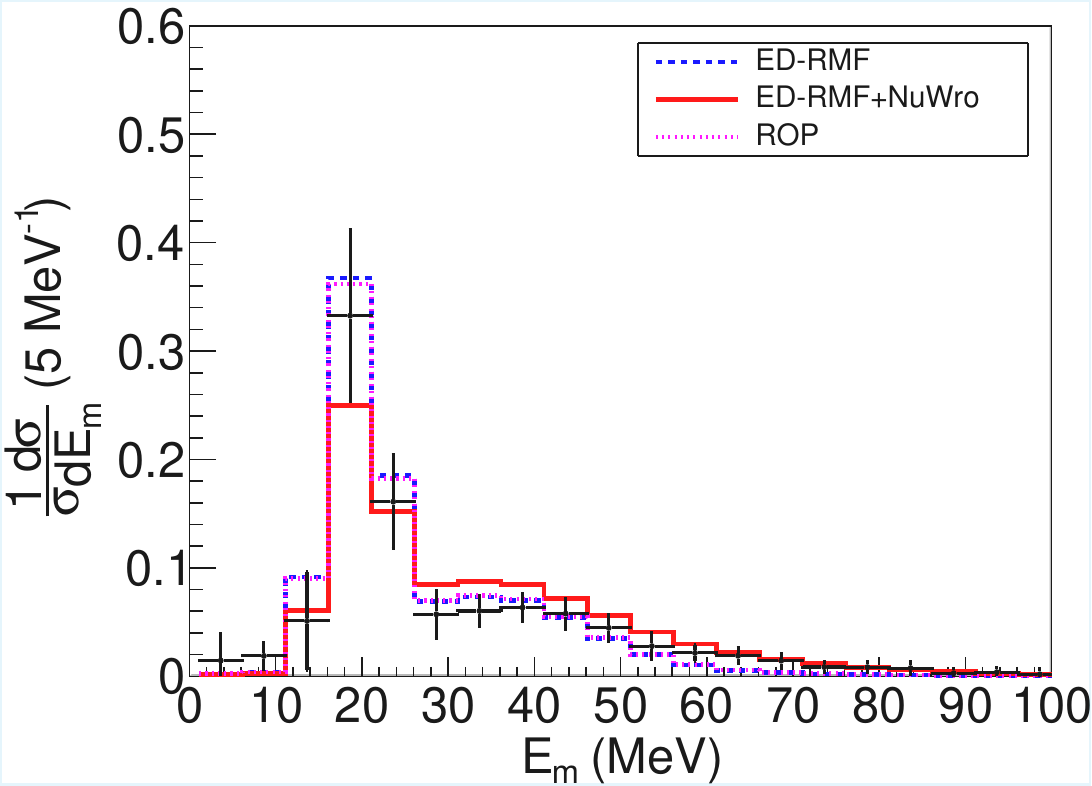}}}%
    		\:
    		\subfloat[]{{\includegraphics[width=0.49\textwidth]{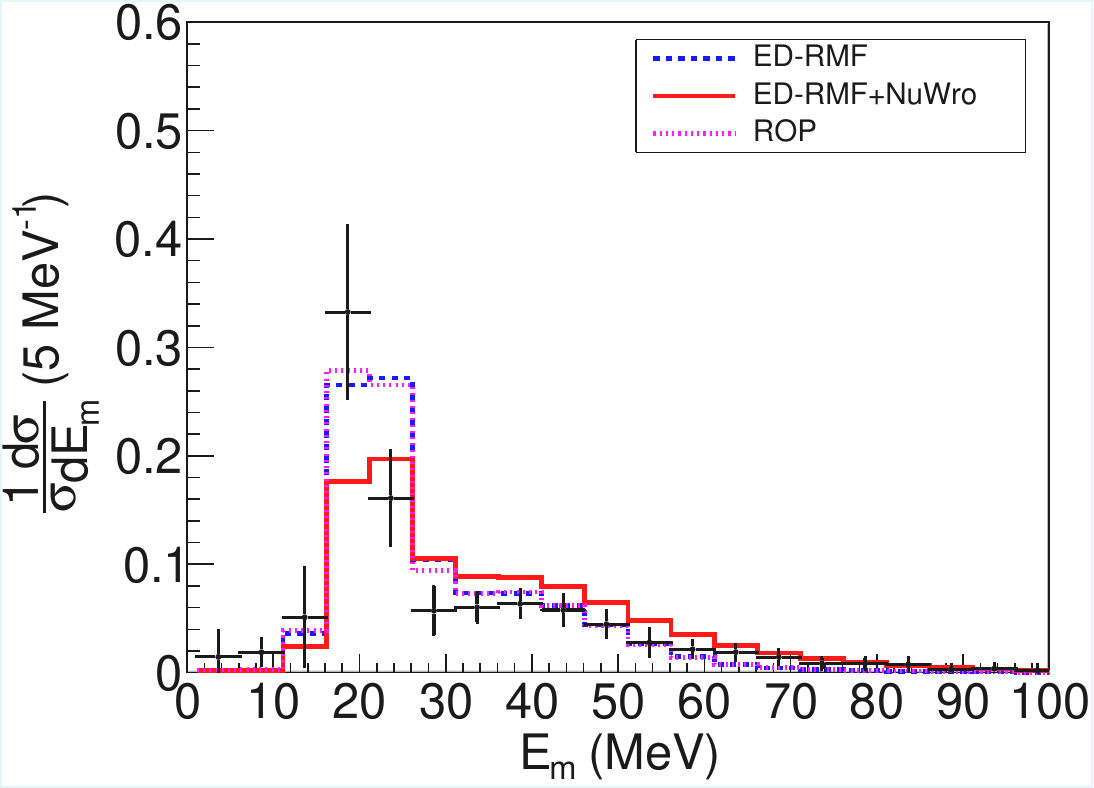}}}%
    	}
    	\caption{\label{fig:smeared} Same than Fig.~\ref{fig:not_smeared} but including a $2.5\%$ energy smearing in the theoretical predictions. }
    \end{figure}
%%%%%%%

The relevance of neutron emission is clearly illustrated in Fig.~\ref{fig:not_smeared} by comparing the ED-RMF + NuWro results with those of the other calculations. The most prominent effect is the migration of events from the region dominated 
by $p$-shells ($E_m < 22$ MeV) to higher values of $E_m$, 
although the other shells are also affected by this mechanism. 
As a result, the ED-RMF + NuWro predictions overestimate the experimental data for $E_m > 30$ MeV in both cases presented in the figure, namely with and without including the recoil of the residual nucleus.

However, predictions from intranuclear cascade models for protons in the kinetic-energy range relevant to KDAR neutrinos must be interpreted with great care. Investigations of hadronic final-state interactions in neutrino–nucleus reactions have found substantial differences among different cascade-model predictions, and they also point to limitations of the semiclassical treatment at such low nucleon energies, where the Compton wavelength becomes comparable to the nuclear size~\cite{PhysRevD.104.053006,PhysRevC.110.054611}.

All the quasielastic contributions, which dominate the results shown in Fig.~\ref{fig:not_smeared}, vanish below the neutron emission threshold of 18.7 MeV, as expected within our RDWIA models. However, the experimental data exhibit nonzero values even at energies below this threshold.

The presence of cross section below threshold may be attributed to nuclear structure effects, correlations, and also to the finite energy resolution of the detector. We simulate the detector smearing with a Gaussian distribution centered at $E_{\textnormal{vis}}$ and with a width $\sigma = 0.025\times E_{\textnormal{vis}}$. This $2.5\%$ energy smearing is quoted in~\cite{mar25} to be a plausible estimate of the minimum energy resolution of the JSNS$^2$ detector. 

The results obtained with this smearing procedure are shown in Fig.~\ref{fig:smeared}. We observe that the region below the threshold is now populated by the theoretical cross sections, and the agreement with the experimental data is improved, especially for the case including the recoil. In the lower panel of Fig.~\ref{fig:smeared}, the position of the peak is no longer visible around $E_m \approx 18.7$ MeV due to the combined effect of neglecting the recoil and the detector smearing. If the recoiling residual nucleus does not produce visible light in the detector, this combined effect could help explain the discrepancies observed between the data and the generator predictions, which does not account for recoil effects in the kinematics, in the region of the peak~\cite{mar25}. In Table~\ref{table:chi2} we show the $\chi^2$ values for all the models presented in Figs.~\ref{fig:not_smeared}-\ref{fig:smeared} using the high statistics region of $E_m=5\textnormal{-}85$ MeV and the full covariance matrix included with the experimental data~\cite{mar25}. For the case without smearing, shown in Fig.~\ref{fig:not_smeared}, there is a large $\chi^2$ improvement in all predictions when the residual nucleus recoil is neglected from the visible energy due to the improvement around the p-shell peak, especially for the ED-RMF and ROP models. The best match of the experimental data is obtained with the ED-RMF+NuWro model with energy smearing and recoil shown in Fig.~\ref{fig:smeared}, with a substantial improvement over the case without smearing due to the migration of events from the peak to adjacent bins, including below the one neutron emission threshold.
        \begingroup
        \setlength{\tabcolsep}{0.6pt}
        \begin{table}[!h]
            \centering
            \begin{tabular}{lcccc}
                \toprule
                & \multicolumn{2}{c}{w/o smearing} & \multicolumn{2}{c}{w/ smearing} \\
                \cmidrule(lr){2-3} \cmidrule(lr){4-5}
                & w/ recoil \,& w/o recoil\,\,&w/ recoil\,&\, w/o recoil\\
                \midrule
                ED-RMF        & 92.5/16 & 25.2/16  & 34.5/16  & 25.6/16  \\
                ED-RMF+NuWro  & 55.2/16 & 30.8/16  & 19.8/16  & 31.4/16  \\
                ROP           & 88.9/16 & 24.5/16  & 32.1/16  & 23.4/16  \\
                \bottomrule
            \end{tabular}
            \caption{\label{table:chi2}Calculated $\chi^2/\mathrm{d.o.f}$ values for each model considered in this work. Only higher statistics region of $E_m = 5–85$ MeV across 16 bins have been considered, as in~\cite{mar25}.}
        \end{table}
        \endgroup

\section{Summary and conclusions}\label{sec:conclusions} %-----------------------------------------%

We analyzed the results of the JSNS$^{2}$ experiment~\cite{mar25}, which 
observes CC interactions of muon neutrinos produced 
via kaon decay at rest on a $^{12}$C target.
In our model, the $\nu_\mu$ interacts with a neutron in $^{12}$C, 
which is converted into a proton via a CC interaction and subsequently 
emitted from the nucleus together with the associated $\mu^-$.

The $^{12}$C ground state is described within a 
relativistic mean-field model. In this framework, the nucleons in 
$^{12}$C fully occupy the lowest single-particle levels, obtained by solving 
a set of coupled Dirac equations that include a potential simulating the 
average interaction of a nucleon with the remaining nucleons~\cite{RING1996193,SHARMA1993377}.
The missing energy, $E_m$, takes discrete values corresponding to the 
single-particle level energies. Consequently, the missing-energy distribution 
$\rho_\kappa\left(E_m\right)$ consists of a sum of Dirac delta functions for all the occupied shells.

We improve this model by constructing a $\rho_\kappa(E_m)$ that reproduces the spectral function of Ankowski et al.~\cite{PhysRevC.110.054612}. Our $\rho_\kappa(E_m)$ parameterization consists of three contributions. The first is a sum of three delta functions accounting for the $p$-shell states. The second describes the $s$-shell contribution, modeled by a Maxwell–Boltzmann distribution. Finally, we include a background term that accounts for the effects of short-range correlations, which generate strength at large missing-energy values.

The final nuclear state, consisting of the emitted proton and the residual $^{11}$C nucleus, is treated using three different approaches. In the first approach (ED-RMF), the proton wave function is distorted by the same real potential employed to describe the $^{12}$C ground state. In the second approach (ROP), the proton wave function is obtained by solving the Dirac equation with a complex optical potential. Finally, the ED-RMF wave function is used as input to the cascade model of NuWro, which simulates the re-scattering of the primary proton with the remaining nucleons within a semiclassical framework. This last modeling approach allows us to identify the role of neutron emission induced by re-scattering processes.

The comparison with the data is affected by an ambiguity in the definition 
of the missing energy, depending on whether the recoil of the residual 
nucleus is taken into account. We therefore present our results for both scenarios, 
namely with and without nuclear recoil. We find that including the recoil effect in our calculations leads to improved agreement with the experimental data, especially in the region of the $p$-shell peak. However, if JSNS$^2$ detector does not collect the light produced by the recoil, the shifted distribution obtained without the recoil could explain the inconsistencies found between the data and the generator predictions, which do not take into account the recoil in the kinematics, in the region of the peak.

The experimental data provide only the shape of the cross section as a 
function of the missing energy, not its absolute normalization. For this reason, 
all our results are normalized to the total strength of the empirical data. 
The ED-RMF and ROP calculations exhibit very similar profiles, 
although the overall strength of the ROP cross section is approximately 27\% 
smaller than that of the ED-RMF result.

The situation is quite different for the ED-RMF + NuWro results. The shape of the distribution is modified, as part of the strength is shifted from the main $p$-shell contribution to larger missing-energy values. This effect arises from neutron emission induced by the re-scattering of the primary proton. Since the neutron energy is not detected by the JSNS$^{2}$ detector, the visible energy $E_{\rm vis}$ is reduced, consequently leading  to a larger reconstructed missing energy (see Eqs.~\eqref{eq:emiss1} and~\eqref{eq:evis}). Including this mechanism leads to poorer agreement with the data, unlike the ED-RMF result. Ref.~\cite{PhysRevD.104.053006} compares proton–$^{12}$C reaction cross-section  with the predictions of different neutrino event generators and finds sizable discrepancies among the generators for $T_p < 50$  MeV, the kinematic region relevant for JSNS$^2$. Nevertheless, even when a generator reproduces the low-$T_p$ reaction cross-section, the observable is inclusive and does not ensure an accurate modeling of particular inelastic final states, such as channels involving neutron emission.

Another effect we have considered is the presence of cross section below the neutron-emission threshold. In our mean-field model, such contributions are forbidden. This discrepancy may arise from nuclear-structure effects not included in mean-field approaches, such as collective nuclear states~\cite{PhysRevC.103.064603} and discrete nuclear excitations leading to emission of photons~\cite{PhysRevD.109.036009,mckean2026}, and/or from the finite energy resolution of the JSNS$^{2}$ detector. We simulate an energy smearing of the measured quantity $E_{\textnormal{vis}}$ with a Gaussian distribution centered at this value and width equal to 2.5$\%$ of $E_{\textnormal{vis}}$. This procedure leads to an improved agreement with the experimental data.

The JSNS$^{2}$ experiment provides important insights into the 
nuclear-structure mechanisms underlying neutrino–nucleus interactions. 
Nevertheless, many questions remain open from both the experimental 
and theoretical viewpoints.

From an experimental perspective, the treatment of nuclear recoil energy 
remains ambiguous. Our results indicate that including recoil yields better agreement with the data than neglecting it. Yet the unfolding procedure of the data relies on a Monte Carlo event generator whose kinematics do not incorporate recoil. If the kinetic energy of the residual nucleus is not effectively detected, the generator would fail to describe the right position of the peak.

From the theoretical side, several issues remain to be addressed. For instance, the importance of low-energy collective excitations requires further investigation. In addition, low-energy $p-^{12}$C reaction cross-section data suggest that the semi-classical description of inelastic FSI in cascade models might be inadequate in this energy regime.

The JSNS$^{2}$ experiment opens new opportunities in neutrino physics 
that deserve further investigation. In particular, measurements of absolute cross sections, beyond shape-only distributions, would be crucial to fully exploit the reach of these data.

\acknowledgments  %-----------------------------------------%
    This work has been partially supported by INFN under Project NUCSYS. R.G.J. and J.G.-M. were supported by project PID2021-127098NA-I00 funded by MCIN/AEI/10.13039/501100011033/FEDER. In addition, R.G.J. was supported by project RYC2022-035203-I funded by MCIN/AEI/10.13039/501100011033/FSE+, UE; and by “Ayudas para Atracción de Investigadores con Alto Potencial–modalidad A” funded by VII PPIT-US. J.G.-M. was also supported by the Fund for Scientific Research Flanders (FWO). V.B. was supported by “Planes Complementarios de I+D+i” program (Grant No. ASFAE/2022/022) by MICIU with funding from the European Union NextGenerationEU and Generalitat Valenciana; by Grant No. PID2023-147458NB-C21 funded by MCIN/AEI/ 10.13039/501100011033 and by the European Union.

\bibliography{nuovo}
\end{document}